\documentclass{mem}
\usepackage{natbib}\usepackage{txfonts}\usepackage{balance}
\usepackage{graphicx}
\usepackage[a4paper,breaklinks,dvipdfm]{hyperref}

\idline{0}{1}

\usepackage{epstopdf}

\begin{document}
\title{Stochasticity effects on derivation of physical parameters of unresolved star clusters}

\author{P. de Meulenaer\inst{1,2} \and D. Narbutis\inst{1} \and T. Mineikis\inst{1,2} \and V. Vansevi\v{c}ius\inst{1,2}}

\institute{Vilnius University Observatory, \v{C}iurlionio 29, Vilnius LT-03100, Lithuania\\\email{philippe.demeulenaer@ff.stud.vu.lt} \and Center for Physical Sciences and Technology, Savanori\c{u} 231, Vilnius LT-02300, Lithuania}

\authorrunning{de Meulenaer et al.}

\titlerunning{Physical parameters of unresolved clusters}

\abstract{We developed a method for a fast modeling of broad-band $UBVRI$ integrated magnitudes of unresolved star clusters and used it to derive their physical parameters (age, mass, and extinction). The method was applied on M33 galaxy cluster sample and consistency of ages and masses derived from unresolved observations with the values derived from resolved stellar photometry was demonstrated. We found that interstellar extinction causes minor age-extinction degeneracy for the studied sample due to a narrow extinction range in M33.
}

\maketitle{}

\section{Introduction}
Star clusters are essential objects to constrain star formation history of their host galaxy. Traditional way to derive physical parameters of unresolved clusters is to compare their integrated broad-band photometry magnitudes to the magnitudes of simple stellar population models. However, the accuracy of parameter derivation is limited due to degeneracy between the parameters and stochastic sampling of stars in clusters.

Recently methods, which take stochasticity into account, for computing integrated cluster magnitudes by using Monte-Carlo sampling of stellar initial mass function, were introduced by \cite{Popescu2010} and \cite{Fouesneau2010}. A large grid of cluster models was built to cover wide ranges of age, mass, and extinction (metallicity assumed to be known). 

Physical parameters of clusters are then derived by comparing observations to the model grid. The first method finds a model which best matches observations, while the second is more accurate but much slower, because it builds probability maps in the age-mass-extinction space by exploring all the nodes of the grid and selects the most probable parameters for a given observation.

We present a new method \citep{de_Meulenaer2013} that also builds probability maps of parameters, but does not require to explore all the nodes of the model grid, hence computation time is reduced significantly. We applied this method on a cluster sample of M33 galaxy studied by \cite{SanRoman2009}, \cite{Ma2012}, and new photometry from \cite{Ma2013}.

\section{Method to derive parameters}
The method is sketched in Fig.\,1. It is fast, because we do not explore all the models of the grid in photometric parameter space (panel a), but restrict to the ones located close to the observed absolute magnitudes of cluster, shown by circle, the size of which depends on photometric errors, $\sigma$.

\begin{figure}[]
\resizebox{\hsize}{!}{\includegraphics[clip=true]{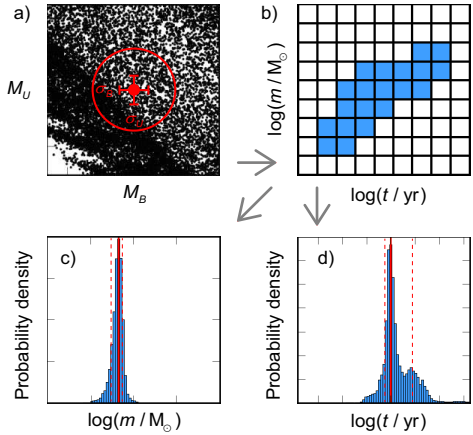}}
\caption{\footnotesize Scheme of the method to derive physical parameters of star cluster; see the text for description.}
\end{figure}

We use Padova isochrones by \cite{Girardi2010} to build cluster models. The model grid (panel b) has fixed metallicity ($Z=0.008$), and contains: 71 steps in age, $\log(t/{\rm yr}) = 6.6 \div 10.1$; 61 in mass, $\log(m/M_\odot) = 2 \div 5$; and 51 in extinction, $E(B-V) = 0 \div 1$. Each node of the grid (panel b) contains 1000 models. 

Panel (a) shows fraction of photometric parameter space (without extinction) in plane of $M_{U}$ vs $M_{B}$, with cluster's observation and error bars. All models located within the 3--$\sigma$ area, which corresponds to nodes shown in panel (b), are selected. Distributions of mass and age shown in panels (c, d) are derived from the selected models. Most probable values are shown by solid vertical lines with confidence intervals (dashed lines).

\cite{de_Meulenaer2013} performed test on a sample of $10^4$ artificial clusters with Gaussian photometric errors of 0.05 mag applied for each passband to quantify accuracy of the method. Fig.\,2 shows, as gray background, age (panels c, e), mass (panels d, f), and extinction (panel b) derived for artificial sample. The two streaks above and below the one-to-one relation (panel c) in the range of $8 \lesssim \log(t/{\rm yr}) \lesssim 9.5$, first reported by \cite{Fouesneau2010}, are a sign of the age-extinction degeneracy.

\begin{figure*}[]
\resizebox{\hsize}{!}{\includegraphics[clip=true]{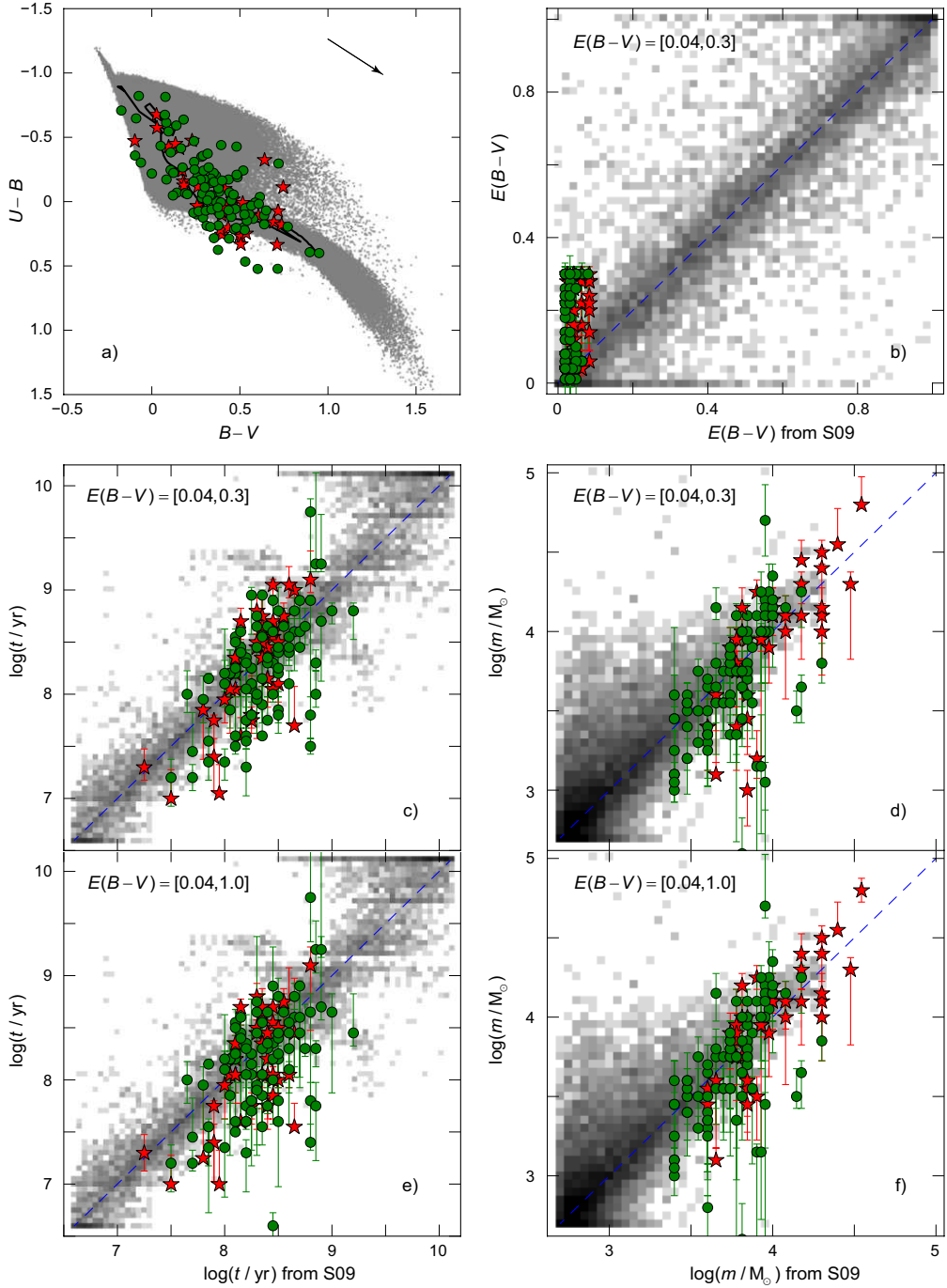}}
\caption{\footnotesize M33 galaxy star clusters common to catalogs of \cite[][S09]{SanRoman2009} and: 1) \cite{Ma2012} (40 stars), 2) \cite{Ma2013} (84 circles). Panel (a): model grid (without extinction) in the background; reddening vector follows LMC extinction law; line traces Padova SSP of $Z=0.008$. Extinction (b), age (c, e), and mass (d, f) derived for clusters using integrated photometry of \cite{Ma2012, Ma2013}, compared to the values based on resolved stellar photometry by S09. Panels (b, c, d) are the case of narrow allowed extinction range, $E(B-V) = [0.04, 0.3]$. Panels (e, f) -- of wide range, $E(B-V) = [0.04, 1.0]$. Background shows results (derived vs true parameters) for artificial cluster sample; see text for description.}
\end{figure*}

\section{Application on M33 clusters}
\cite{SanRoman2009} derived age, mass, and extinction of 161 clusters in M33 galaxy using {\it HST} by isochrone fit to CMDs of their resolved stellar photometry. \cite{Ma2012} provided $UBVRI$ broad-band integrated photometry for 40 of them. Recently, \cite{Ma2013} extended sample by 101 additional clusters and here we present results following study by \cite{de_Meulenaer2013}. Sample is presented in Fig.\,2a; note that $B$ band magnitudes of clusters were corrected by 0.1 mag.

We adopted the M33 distance modulus of $(m-M)_{0} = 24.54$ \citep{McConnachie2005}. Extinction range of the M33 galaxy is assumed to be between the foreground Galactic line-of-sight extinction of $E(B-V)$ = 0.04 \citep{Schlegel1998} and 0.3, derived from supergiants \citep{U2009} and HII regions \citep{Rosolowsky2008}. Interstellar extinction law of LMC \citep{Gordon2003} is used to redden the models. Note, that extinction range for clusters derived by \cite{SanRoman2009} is more narrow (Fig.\,2b) than expected in M33.

The M33 cluster parameters were derived using two different allowed extinction ranges for the model grid: 1) narrow, $E(B-V)$ = [0.04, 0.30], and 2) wide, $E(B-V)$ = [0.04, 1.0], as one used for the M31 case by \cite{de_Meulenaer2013}. Panels (b, c, d) of Fig.\,2 show the extinction, age and mass derived for the narrow extinction range. Panels (e, f) show parameters derived with the wide range. Our results, compared to the values based on isochrone fit to CMDs by \cite{SanRoman2009}, show consistency.

Differences are relatively small between the two kinds of parameter solutions using narrow and wide extinction ranges of the model grid; panels: (c) vs (e), (d) vs (f). This suggests that the age-extinction degeneracy plays a minor role in this M33 cluster sample. Conversely, \cite{de_Meulenaer2013} showed that degeneracy plays a major role in the M31 galaxy preventing accurate parameter derivation for a large fraction of clusters.

\section{Conclusions}
The presented method takes into account stochastic sampling of stars in unresolved star clusters when deriving their physical parameters (age, mass, and extinction) using broad-band photometry. The M33 galaxy star cluster sample studied by \cite{SanRoman2009} using resolved stellar photometry (isochrone fit to CMDs), was analyzed based on the $UBVRI$ integrated photometry from \cite{Ma2012,Ma2013}, and shows consistent ages and masses.

By studying the cluster sample successively in narrow and wide allowed extinction ranges, we demonstrate that derived cluster parameters indicate low sensitivity to the age-extinction degeneracy, contrarily to what is the case for the M31 galaxy star clusters \citep{de_Meulenaer2013}.

\begin{acknowledgements}
This research was funded by a grant (No. MIP-102/2011) from the Research Council of Lithuania.
\end{acknowledgements}

\end{document}